\documentclass[apj,twocolumn]{openjournal}
\usepackage{amsmath}
\usepackage{booktabs}
\usepackage{multirow}
\usepackage{color}
\usepackage{soul}

\usepackage[dvipsnames]{xcolor} %added later for color

\usepackage[breaklinks,colorlinks,citecolor=blue,urlcolor=blue]{hyperref}

\defcitealias{Shen2018}{S18}

\renewcommand{\arraystretch}{1.2}  
\usepackage{listings}
\usepackage{color}
\definecolor{dkgreen}{rgb}{0,0.6,0}
\definecolor{gray}{rgb}{0.5,0.5,0.5}
\definecolor{mauve}{rgb}{0.58,0,0.82}
\definecolor{golden}{rgb}{0.86,0.65,0.01}
\usepackage{orcidlink}

\begin{document}

%\preprint{}

\title{How to use {\it Gaia} parallaxes for stars with poor astrometric fits}

%\author{\vspace{-30pt}Kareem El-Badry$^{1}$}
\author{\vspace{-30pt}Kareem El-Badry\,\orcidlink{0000-0002-6871-1752}$^{1}$}

\affiliation{$^1$Department of Astronomy, California Institute of Technology, 1200 E. California Blvd., Pasadena, CA 91125, USA}
\email{Corresponding author: kelbadry@caltech.edu}

\begin{abstract}
\textit{Gaia}  parallax measurements for stars with poor astrometric fits -- as evidenced by high renormalized unit weight error (\texttt{RUWE}) -- are often assumed to be unreliable, but the extent and nature of their biases remain poorly quantified. High \texttt{RUWE} is usually a consequence of binarity or higher-order multiplicity, so the parallaxes of sources with high \texttt{RUWE} are often of greatest astrophysical interest. Using realistic simulations of {\it Gaia} epoch astrometry, we show that the parallax uncertainties of sources with elevated \texttt{RUWE} are underestimated by a factor that ranges from 1 to 4 and can be robustly predicted from observables. We derive an empirical prescription to inflate reported uncertainties based on a simple analytic function of \texttt{RUWE}, apparent magnitude, and parallax. We validate the correction using (a) single-star solutions for {\it Gaia} sources with known orbital solutions and (b) wide binaries containing one component with elevated \texttt{RUWE}. The same uncertainty corrections are expected to perform well in DR4 and DR5. Our results demonstrate that {\it Gaia} parallaxes for high-\texttt{RUWE} sources can still yield robust distance estimates if uncertainties are appropriately inflated, enabling distance constraints for triples, binaries with periods too long or too short to be fit astrometrically, and sources blended with neighboring sources. 
\keywords{astrometry -- parallaxes -- binaries}
\end{abstract}

\maketitle

\section{Introduction}
\label{sec:intro}
Data from the {\it Gaia} mission has enabled parallax measurements -- and thus, distance constraints -- for about 2 billion stars \citep{GaiaCollaboration2016, GaiaCollaboration2023b}. For astrometrically well-behaved sources that are effectively single, motion on the plane of the sky can be fully described by a linear model with five free parameters: two positional coordinates, two proper motions, and parallax. The parallax is often the quantity of greatest interest, because it constrains the distance to the star \citep[e.g.][]{Bailer-Jones2015, Luri2018}. 

Some {\it Gaia} sources have spurious parallax measurements. In this context, ``spurious'' does not necessarily mean that the parallax contains no useful information, but rather that the difference between the reported parallax and its true value significantly exceeds the reported uncertainty. The existence of such sources is manifest, for example, in the prevalence of sources with significantly negative parallaxes \citep[e.g.][]{Rybizki2022}. 

Most spurious {\it Gaia} astrometric solutions are due to binaries and higher-order multiples. Binaries with detectable astrometric ``wobble'' on the sky -- typically systems with orbital periods of months to years -- lead to complicated motion of the photocenter that cannot be described by a five-parameter astrometric solution. Binaries with much longer orbital periods of centuries to millennia can similarly lead to spurious astrometric solutions, because blended light from a marginally resolved companion perturbs astrometric measurements \citep[e.g.][]{Lindegren2022}. Physically unassociated close neighbors that are in the foreground or background of a source can have the same effect. 

Several parameters quantify the goodness of fit of {\it Gaia} five-parameter solutions. The most useful of these is  \texttt{RUWE} (``renormalized unit weight error'') which approximately represents the reduced chi-squared, $\chi_\nu^2$, of the five-parameter astrometric solution \citep{Lindegren_2018_RUWE}. Several previous works have explored the utility of \texttt{RUWE} as diagnostic of spurious solutions and binary companions. \citet{Lindegren_2018_RUWE} proposed that for {\it Gaia} DR2, a cut of \texttt{RUWE} $< 1.4$ would retain most ``good'' solutions while filtering out most spurious ones.\footnote{For DR3, a threshold of \texttt{RUWE} $<1.25$ has been proposed to be more appropriate, though the exact threshold varies with sky position \citep{Penoyre2022}.} \citet{Belokurov2020} showed that, at least for nearby sources, \texttt{RUWE} $> 1.4$ primarily selects sources in regions of the color-magnitude diagram (CMD) where binaries are expected to be overrepresented. \citet{Penoyre2020} developed a framework to predict \texttt{RUWE} and other astrometric observables for binaries; \citet{Penoyre2022}, \citet{Penoyre2022b}, and \citet{Korol2022} used this framework to infer properties of the local binary population from the observed \texttt{RUWE} distribution. Most recently, \citet{Castro-Ginard2024} developed a model to quantify the selection function of \texttt{RUWE}-based binary samples, while \citet{Kiefer2024} used \texttt{RUWE} to select candidate exoplanet hosts. 

It is frequently assumed in the literature that parallaxes for source with large \texttt{RUWE} (most commonly, \texttt{RUWE} $> 1.4$) are unreliable. However, it has not been quantified {\it how} unreliable, and under what conditions such parallaxes still contain useful information. One might expect, for example, that larger \texttt{RUWE} values result in less reliable parallaxes, or that nearby sources with large \texttt{RUWE} are less affected than distant ones with the same \texttt{RUWE}. Sources with elevated  \texttt{RUWE} are often of particular astrophysical interest: for example, binaries that receive spectroscopic solutions but not astrometric orbital solutions are likely to have large \texttt{RUWE} \citep[e.g.][]{Thompson2019, El-Badry2021_hr6819,
 Yamaguchi2024,Wang2024}, and distance is often the key uncertainty in measuring the physical properties of these systems. 

Poor astrometric solutions as quantified by \texttt{RUWE} are quite common. This is unsurprising, since an order unity fraction of all {\it Gaia} sources are binaries. For example, 17\% of {\it Gaia} DR3 sources in the {\it Gaia} Catalog of Nearby Stars \citep{Smart2021} have \texttt{RUWE} $>1.4$, and 22\% have \texttt{RUWE} $> 1.25$. It is thus desirable to understand how to interpret the parallaxes and uncertainties of such sources. In this work, we show that while sources with elevated \texttt{RUWE} have underestimated uncertainties, the magnitude of the difference between the best-fit and true parallax can be accurately predicted from observables. The parallaxes of such sources can thus in most cases still yield useful distance constraints. 

The rest of the paper is organized as follows. In Section~\ref{sec:sims}, we describe simulations of {\it Gaia} epoch astrometry and astrometric model fitting. Section~\ref{sec:experiments} summarizes a series of controlled experiments designed to estimate the parallax bias that results from unmodeled noise in the epoch astrometry. In Section~\ref{sec:fittingfunc}, we present a simple analytic fitting function that captures the trends revealed in our simulations and can be used to inflate published {\it Gaia} parallax uncertainties for sources with large \texttt{RUWE}. We validate this function using binaries with astrometric orbits and resolved wide binaries in Section~\ref{sec:validation}. We discuss our results and conclude in Section~\ref{sec:conclusion}.

\section{Description of the simulations}
\label{sec:sims}

We simulate astrometric orbits for binaries using the \texttt{gaiamock}\footnote{\href{https://github.com/kareemelbadry/gaiamock}{https://github.com/kareemelbadry/gaiamock}} model, which is described in detail by \citet{Elbadry_2024}. \texttt{gaiamock} generates realistic 1D epoch astrometry for binaries or single stars from the {\it Gaia} scanning law, using an empirical noise model based on the residuals of well-behaved astrometric sources. It then fits the epoch astrometry with an astrometric model; here, we consider only the five-parameter single-star model. 

Following the {\it Gaia} convention, \texttt{gaiamock} inflates the astrometric uncertainties of sources with poor fits. This is accomplished by inflating all the epoch astrometry uncertainties, and the uncertainties of the astrometric parameters, by a constant $c=\sqrt{\chi^{2}/\left[\nu\left(1-\frac{2}{9\nu}\right)^{3}\right]}\approx\sqrt{\chi_{\nu}^{2}}$. Here $\nu$ is the number of astrometric epoch measurements minus five and is typically $\approx 300$ in DR3.  As we will show, this inflation is generally not sufficient, because the five-parameter astrometric model is mismatched to the data. \texttt{RUWE} for the five-parameter solution is calculated as 

\begin{align}
    \label{eq:ruwe}
    \texttt{RUWE} = \sqrt{\chi_\nu ^2},
\end{align}
meaning that the constant factor by which the published parallax uncertainties are inflated is approximately equal to \texttt{RUWE}. As we show below, $\texttt{RUWE}$ calculated from Equation~\ref{eq:ruwe} is directly comparable to \texttt{RUWE} reported in the {\it Gaia} archive -- there is no need for renormalization, because the epoch uncertainties used by \texttt{gaiamock} are estimated empirically from the residuals of well-behaved sources, not from the formal errors. 

\subsection{Two-parameter solutions}
Another quantity required for constructing a realistic mock astrometric catalog is $\sigma_{\rm 5d,\,max}$ (reported as \texttt{astrometric\_sigma5d\_max} in the {\it Gaia} archive), which is calculated as the square root of the largest singular value of the scaled $5\times 5$ covariance matrix of the astrometric parameters. The ``scaling'' is as described in \citet{Lindegren2018b}, with $T$ taken to be the nominal DR3 baseline of 2.76383 yr, and singular values calculated from the covariance matrix {\it after} uncertainty inflation. Roughly speaking, $\sigma_{{\rm 5d,\,max}}$ represents the most uncertain astrometric parameter or linear combination of said parameters. Sources with $\sigma_{{\rm 5d,\,max}}>10^{0.2\,{\rm max}\left(6-G,0,G-18\right)}$ do not receive five-parameter solutions and are instead published with 2-parameter solutions. The main effect of this cut is that it removes sources with very large formal astrometric uncertainties, which are mainly marginally resolved wide binaries with similar-brightness components \citep{Tokovinin2023}. The cut also results in a maximum \texttt{RUWE} of $\sim 50$ for bright sources (and lower for fainter sources), since \texttt{RUWE} is not calculated for two-parameter solutions. 

\subsection{Validating \texttt{RUWE} predictions from \texttt{gaiamock} }
\label{sec:validation_nss}
\begin{figure}
    \centering
    \includegraphics[width=\columnwidth]{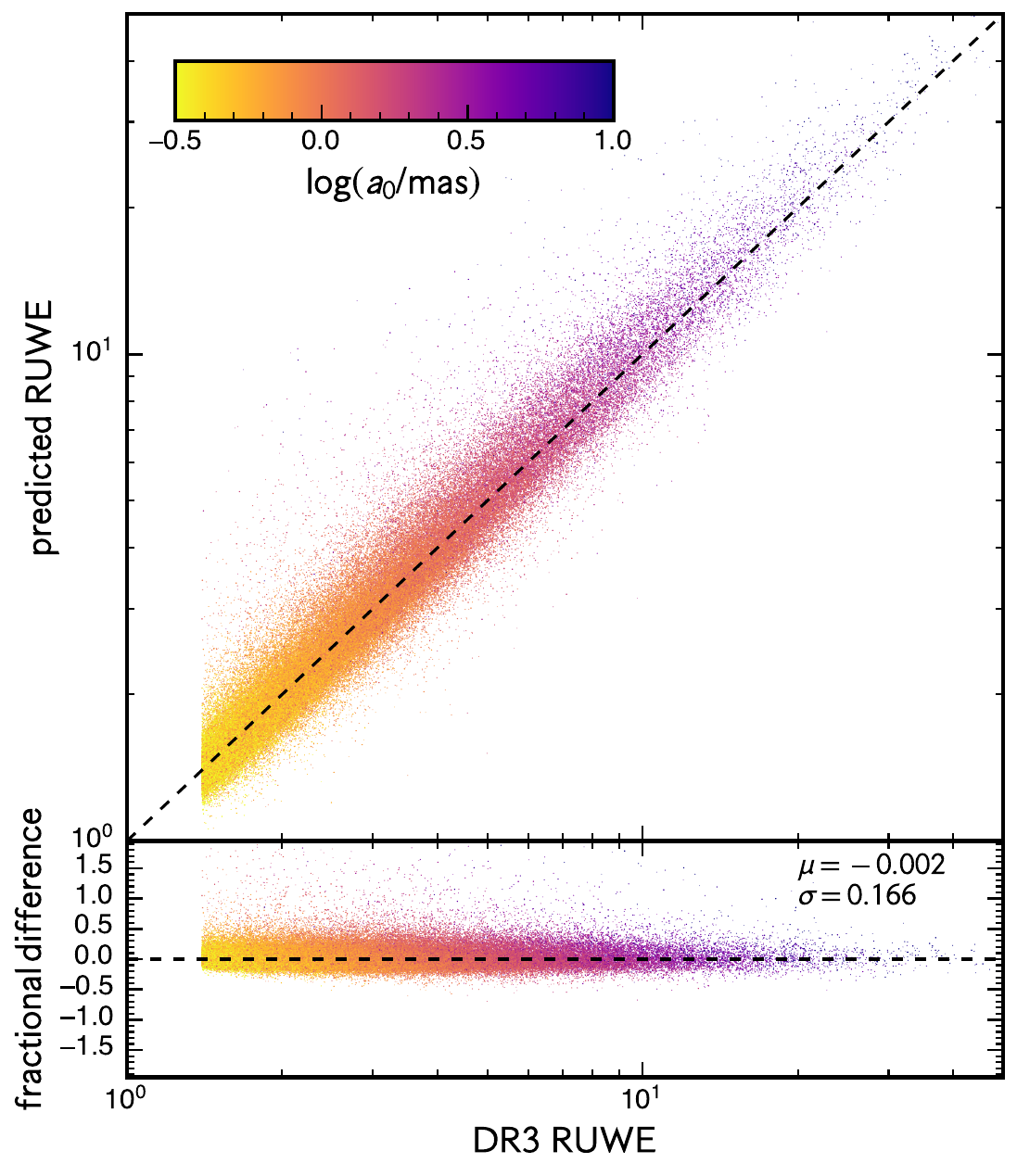}
    \caption{Predicted versus true \texttt{RUWE} for binaries with astrometric orbital solutions published in DR3. X-axis shows \texttt{RUWE} reported in DR3 for the sources' five-parameter solutions; Y-axis shows \texttt{RUWE} predicted by \texttt{gaiamock} from the binary's known orbit as measured by the 12-parameter solution. Sources are colored by $a_0$, the angular semimajor axis of the photocenter ellipse, which is correlated with \texttt{RUWE}. The good agreement -- with negligible bias and scatter of 17\% -- implies that \texttt{gaiamock} can robustly predict \texttt{RUWE}. }
    \label{fig:nss}
\end{figure}

To verify that \texttt{gaiamock} accurately predicts \texttt{RUWE}, we begin by using the code to predict \texttt{RUWE} of the five-parameter solutions of binaries that received full 12-parameter astrometric orbits in DR3 \citep{Halbwachs2023, GaiaCollaboration2023}. Of the 168,065 astrometric orbits published in DR3, we retain 154,778 good orbits satisfying \texttt{goodness\_of\_fit} $< 5$ for $G > 13$ and \texttt{goodness\_of\_fit} $< 10$ for $G \leq 13$ \citep[see][]{El-Badry2023}. We predict their DR3 epoch astrometry and five-parameter solutions using \texttt{gaiamock}; the corresponding \texttt{RUWE} values are shown in Figure~\ref{fig:nss}.

\texttt{gaiamock} predicts \texttt{RUWE} for DR3 binaries quite accurately, with negligible systemic offset and scatter of 16.6\%. \texttt{RUWE} is strongly correlated with $a_0$, which is not surprising, because it is essentially a measure of the size of typical residuals for the five-parameter solution (which scales with $a_0$) in units of the per-epoch astrometric uncertainties. However, predicting \texttt{RUWE} with \texttt{gaiamock} results in a tighter correlation than any simple scaling with $a_0$. The tight correlation shown in Figure~\ref{fig:nss} demonstrates that \texttt{gaiamock} predicts \texttt{RUWE} values for binaries robustly. At low \texttt{RUWE}, there is a slight bias toward higher predicted than observed \texttt{RUWE}, perhaps hinting at imperfectly modeled systematics in the low-\texttt{RUWE} regime. This, however, only amounts to a median ruwe difference of 0.02 for sources with \texttt{RUWE} $<1.5$. 

\section{Results} 
\label{sec:experiments}
We use \texttt{gaiamock} to simulate orbits of samples of binaries at fixed apparent magnitude and distance. For each such experiment, we draw $10,000$ orbital periods from the lognormal distribution observed for solar-type binaries \citep{Raghavan2010} and mass ratios from a uniform distribution. We assume $1\,M_{\odot}$ primaries and assign $G-$band flux ratios assuming both components are at the zero-age main sequence, assign the binaries random orientations and phases, and draw eccentricities from a Rayleigh distribution with $\sigma_e = 0.34$ \citep{Wu2025}. Our results are not sensitive to any of these choices, because we are ultimately investigating parallax bias at fixed \texttt{RUWE}, not at fixed binary parameters. 

\begin{figure*}
    \centering
    \includegraphics[width=\textwidth]{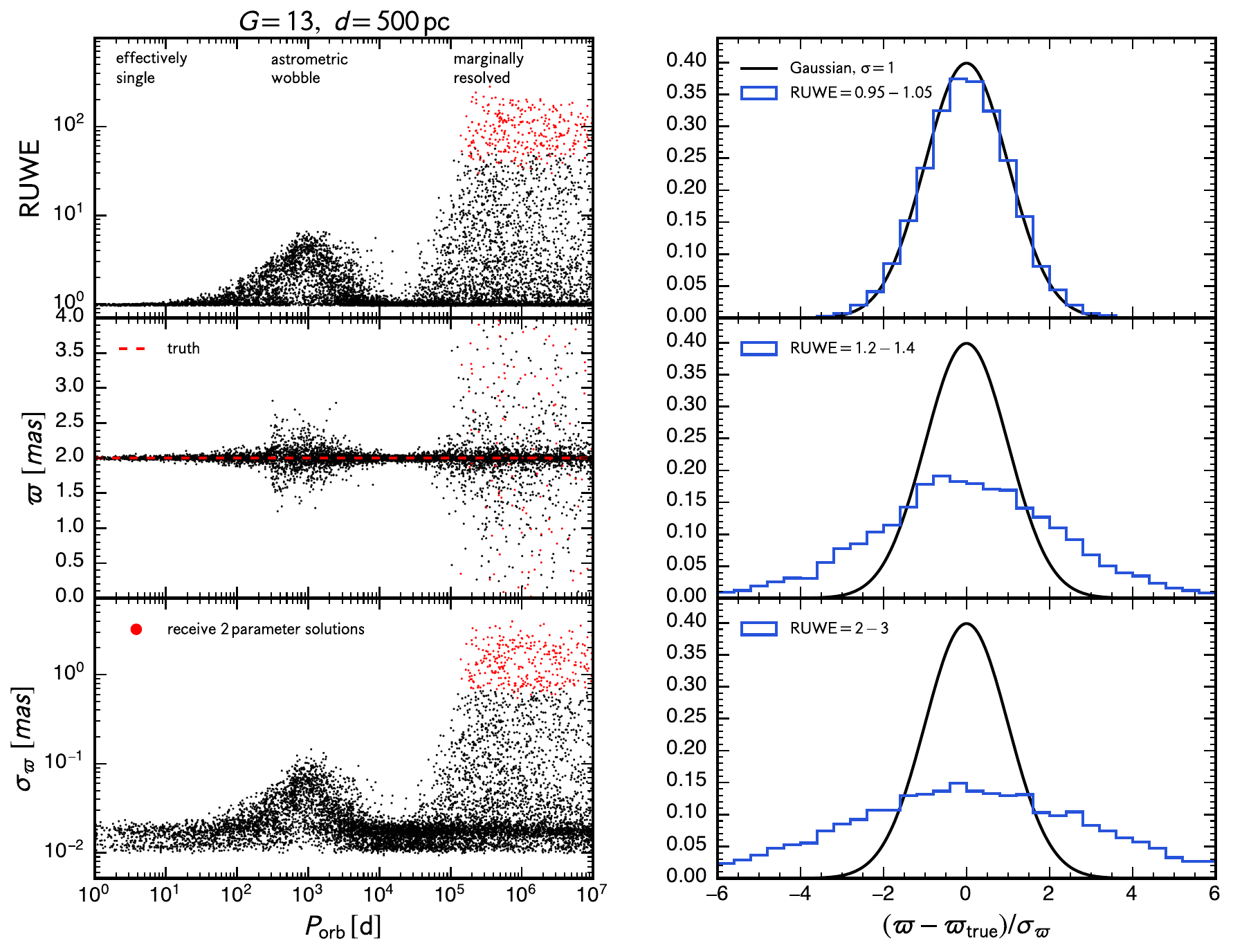}
    \caption{Left: \texttt{RUWE}, parallax, and formal parallax uncertainty for simulated binaries with $G=13$ and $d=500$\,pc. We assume a realistic distribution of orbital periods, mass ratios, flux ratios, and sky positions. Short-period binaries ($P_{\rm orb} \lesssim 100$\,d are effectively single, as their orbits are too small to lead to significant photocenter wobble. Binaries with periods of $100-5000$\,d exhibit elevated \texttt{RUWE} and large scatter in best-fit parallax around the true value due to astrometric wobble of the unresolved photocenter. Finally, binaries with $P_{\rm orb} \gtrsim 10^5$ d are marginally resolved, such that {\it Gaia} does not observe the photocenter, again leading to large \texttt{RUWE}. {\it Gaia's} formal parallax uncertainty (bottom panel) is elevated in the same regions where there is scatter in the best-fit parallaxes, but the correlation is not perfect. Most notably, systems with periods near 1 year exhibit large scatter in best-fit parallax but only modest \texttt{RUWE} and $\sigma_\varpi$. Right: distribution of parallax bias in three \texttt{RUWE} bins, normalized by the formal parallax uncertainty. A distribution wider than a $\sigma=1$ Gaussian (black line) indicates underestimated parallax uncertainties. We measure uncertainty inflation factors (Figure~\ref{fig:fitting_func}) from the width of these distributions.    }
    \label{fig:vs_period}
\end{figure*}

Figure~\ref{fig:vs_period} shows the results of these simulations for a single bin of apparent magnitude and distance. In the left panel, we show \texttt{RUWE} (top), best-fit parallax (middle), and reported parallax uncertainty (bottom)  for orbital periods ranging from 1 to $10^7$ d. 

At short periods ($P_{\rm orb}\lesssim 100$\,d for the distance shown in Figure~\ref{fig:vs_period}), binaries have accurate parallaxes and uncertainties and \texttt{RUWE} $\approx 1$. This reflects the fact that these binaries have photocenter orbits that are small compared to the {\it Gaia} DR3 measurement uncertainties. At periods ranging from $100-1000$ d, \texttt{RUWE} rises with increasing $P_{\rm orb}$, because longer-period binaries have larger photocenter orbits that lead to larger deviations from single-star motion. Some binaries in this period range still have low \texttt{RUWE}; these are systems with mass ratios near 0 or 1.

The formal parallax uncertainties -- which are inflated by a factor almost identical to \texttt{RUWE} -- rise in tandem with \texttt{RUWE}. The typical parallax bias also increases, as evidenced by increased scatter in the second panel of Figure~\ref{fig:vs_period}. The scatter is generally largest at period ranges where \texttt{RUWE} is also large, but trends in \texttt{RUWE} and parallax bias with period are not identical. Most notably, there is a sharp increase in parallax bias at $P_{\rm orb} \approx 1$ year that is not matched by a spike in \texttt{RUWE}. As has been demonstrated elsewhere \citep[e.g.][]{Penoyre2020}, this feature is a result of orbital motion being mis-absorbed into the parallax. However, only a small fraction of all binaries fall in the relevant 300-500 d period range. 

All three quantities shown in Figure~\ref{fig:vs_period} drop at periods beyond 1000 d, where observations in {\it Gaia} DR3 have been taken over less than a full orbit and most of the orbital motion can be absorbed into the apparent proper motion. Binaries with $P_{\rm orb} = 10^4-10^5$ d thus receive unproblematic five-parameter solutions with small \texttt{RUWE} and little parallax bias. However, at periods longer than $10^5$ d, all three quantities begin to rise again. This is a result of the binaries' being marginally resolved, such that the position measured by the {\it Gaia} image parameter determination pipeline depends on the scan angle and traces neither the photocenter nor the location of either binary component \citep[e.g.][]{Lindegren2022}. Binaries with flux ratios near 1 are most strongly affected, while those with faint secondaries retain small \texttt{RUWE} and unbiased parallaxes. The largest predicted \texttt{RUWE} values occur in this period range and are a result of marginally resolved pairs with flux ratios near unity. The most extreme cases  (red points) fail the cut on $\sigma_{\rm 5d,\, max}$ and only receive 2-parameter solutions. 

Given a set of simulated sources, we estimate their {\it true} parallax uncertainty (i.e., the expected magnitude of the bias) from the distribution of measured parallaxes. The right panels of Figure~\ref{fig:vs_period} show that for any fixed apparent magnitude (corresponding to fixed along-scan uncertainty $\sigma_\eta$) and distance, the distribution of best-fit parallaxes is close to Gaussian, with the width of this distribution corresponding to the parallax bias. If the formal parallax uncertainties are accurate,  $\left(\varpi-\varpi_{{\rm true}}\right)/\sigma_{\varpi}$ should follow a normal distribution $\mathcal{N}(0,1)$, as in the top panel for sources with $\texttt{RUWE}\approx 1$. For $\texttt{RUWE} > 1$, the distribution broadens, and its width becomes diagnostic of the parallax bias. 

\subsection{Dependence of parallax bias on distance and apparent magnitude}
\label{sec:dependence}

We perform simulations similar to those shown in Figure~\ref{fig:vs_period} for a range of apparent magnitudes, $G=6-20$ in steps of 1 mag, and a range of distances, with 50 logarithmically spaced values between 30 pc and 10 kpc. In each case, we quantify the width of the parallax bias distribution as half the difference between its 16th and 84th percentiles. 

The parallax bias $\delta_\varpi$ (i.e., the ``true'' parallax uncertainty)  is correlated with $\sigma_{\varpi}$, the formal parallax uncertainty. This is not surprising, since $\sigma_{\varpi}$ contains information about both the goodness of fit of the five-parameter solution and the number of observations used in calculating it. We therefore quantify the parallax bias via an uncertainty inflation factor, $f$, which is defined as
\begin{equation}
    \label{eq:bias}
    f=\frac{\delta_{\varpi}}{\sigma_{\varpi}},
\end{equation}
such that $f=1$ corresponds to accurate uncertainties. 
$\delta_{\varpi}$ is defined such that parallax measurements of sources with true parallax $\varpi$ follow a normal distribution $\mathcal{N}(\varpi, \delta_\varpi ^2)$. 

The results are shown in the top panels of Figure~\ref{fig:fitting_func} for selected apparent magnitudes. We find that the shape of $f(\texttt{RUWE})$ depends only weakly on apparent magnitude and distance. In all cases, $f\approx 1$ at $\texttt{RUWE}=1$, increases over $1<\texttt{RUWE}<2$, and plateaus at $\texttt{RUWE}>3$. The asymptotic value of $f$ at large \texttt{RUWE} is $\approx 3$ on average, with a weak dependence on apparent magnitude and parallax. The increase at small \texttt{RUWE} is steep: at \texttt{RUWE} $=1.4$, parallax uncertainties are already underestimated by a factor of 2 on average. The range of \texttt{RUWE} values that occur is smaller at fainter magnitudes, where the per-epoch along-scan uncertainties $\sigma_\eta$ are smaller. 

\subsection{Analytic fitting function}
\label{sec:fittingfunc}

\begin{figure*}
    \centering
    \includegraphics[width=\textwidth]{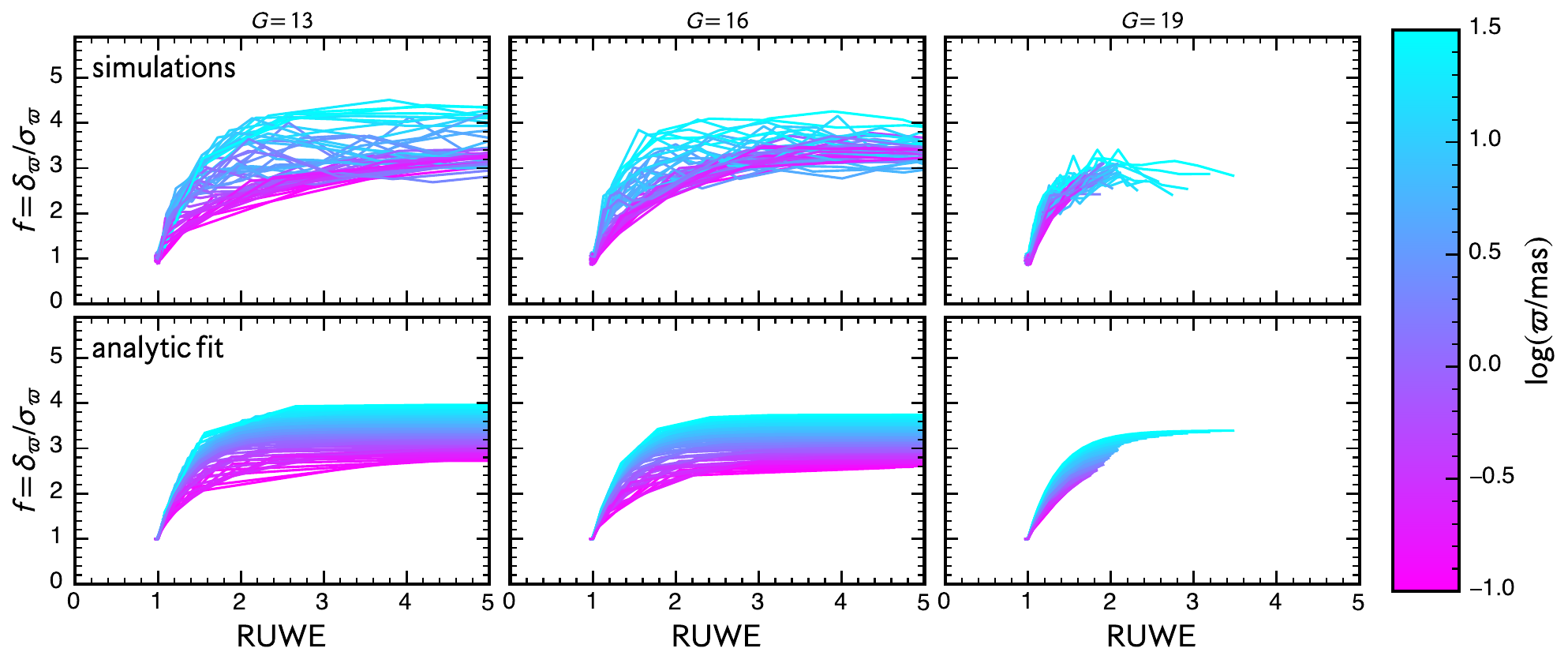}
    \caption{Uncertainty inflation factor to correct published parallax uncertainties for sources with large \texttt{RUWE}. Top panels show results of simulations; bottom panels show our analytic fit (Equation~\ref{eq:f}). For sources with \texttt{RUWE} $\gtrsim 2$, the published uncertainties are generally underestimated by a factor of $\sim 3$. The underestimate is slightly more severe for sources with large parallaxes.    }
    \label{fig:fitting_func}
\end{figure*}

Motivated by the simulation results shown in the top panel of Figure~\ref{fig:fitting_func}, we parameterize the parallax uncertainty inflation factor $f$ as follows:

\begin{equation}
    \label{eq:f}
    f=\begin{cases}
1 & \texttt{RUWE}\leq1\\
1+\left(f_{{\rm max}}-1\right)\left(1-e^{-\alpha\left(\texttt{RUWE}-1\right)}\right), & \texttt{RUWE}>1
\end{cases}.
\end{equation}
This function rises from $f=1$ at $\texttt{RUWE}=1$ to $f\to f_{\rm max}$ at large $\texttt{RUWE}$; the parameter $\alpha$ controls the steepness of the transition. Because our simulations reveal a dependence of the asymptotic inflation factor at large $\texttt{RUWE}$ on parallax and apparent magnitude, we parameterize $f_{\rm max}$ as a power law function of $\varpi$ and $\sigma_{\eta}$:

\begin{equation}
    \label{eq:fmax}
    f_{{\rm max}}=f_{0}\left(\frac{\varpi}{10\,{\rm mas}}\right)^{\beta}\left(\frac{\sigma_{\eta}\left(G\right)}{0.1\,{\rm mas}}\right)^{\gamma}.
\end{equation}

We fit the results of the full simulation grid simultaneously, finding best-fit parameters of $\alpha = 2.77$, $f_0=3.73$, $\beta=0.065$, and $\gamma=-0.056$. In the bottom panels of Figure~\ref{fig:fitting_func}, we plot the corresponding predicted $f$ values. Here $\sigma_{\eta}(G)$ is the empirical ``per-CCD'' along-scan measurement uncertainty shown in Figure 1 of \citet{Elbadry_2024}, which is taken from \citet{Holl2023}.

The fitting function captures most of the structure visible in the data. We use parallax uncertainty inflation factors predicted by Equation~\ref{eq:f} in the remainder of the paper. 

\section{Validation}
\label{sec:validation}

We validate the parallax uncertainty inflation prescription from Equation~\ref{eq:f} using two independent approaches. 

\subsection{Sources with astrometric orbital solutions}
\label{sec:nss_validation}
We first consider the five-parameter astrometric solutions of sources that also received a 12-parameter orbital solution in DR3; i.e., the same sources shown in Figure~\ref{fig:nss}. These sources all have \texttt{RUWE} $>1.4$, and their five-parameter solutions have much larger astrometric uncertainties than their 12-parameter solutions. We can thus consider the 12-parameter solutions as effectively ground truth, allowing us to gauge the true uncertainties of the five-parameter solutions and the effectiveness of the uncertainty inflation. 

\begin{figure*}
    \centering
    \includegraphics[width=\textwidth]{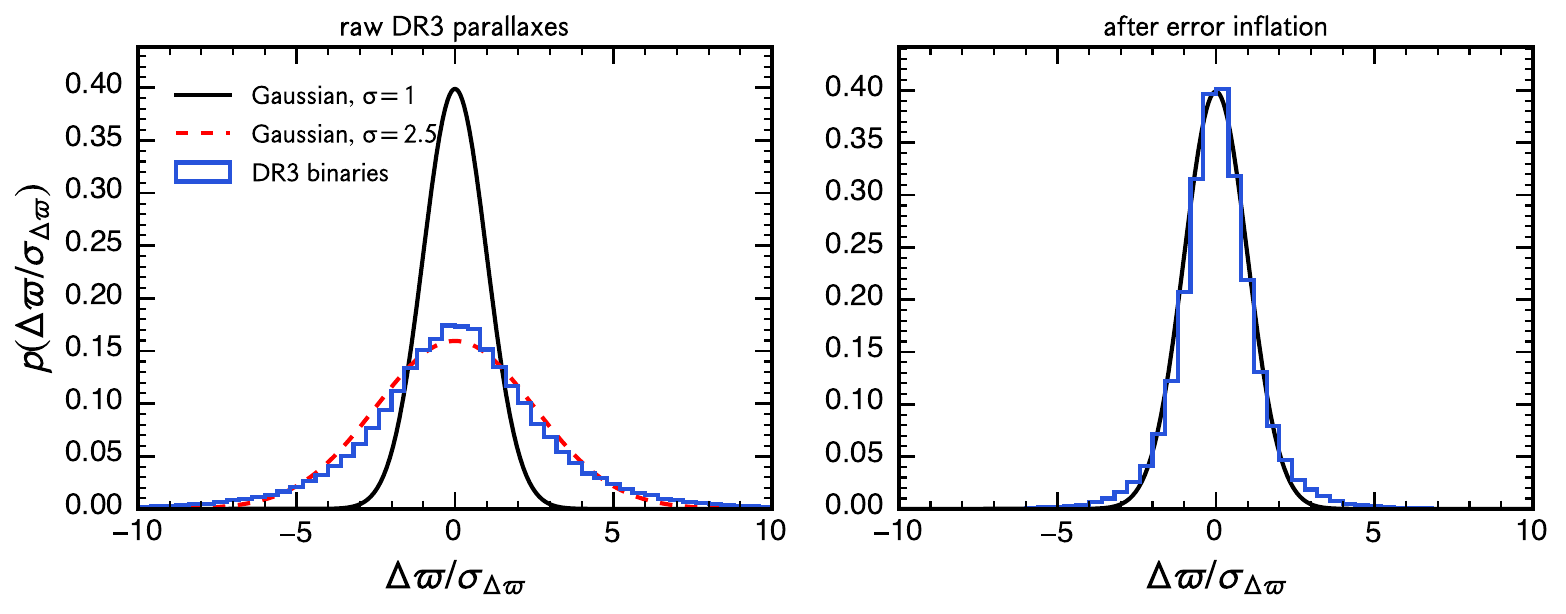}
    \caption{Uncertainty-normalized parallax difference between five-parameter and 12-parameter astrometric solutions for {\it Gaia} sources  published in DR3 with good 12-parameter solutions. If the parallax uncertainties were accurate, the distribution should be Gaussian with $\sigma=1$. Left panel shows that this is not the case when the published parallax uncertainties are used; the distribution is closer to a Gaussian with $\sigma=2.5$, implying that the uncertainties are underestimated by a factor of 2.5 on average. Right panel shows the distribution after the uncertainties of the five-parameter solutions are inflated according to Equation~\ref{eq:f}. The distribution now resembles a $\sigma=1$ Gaussian, implying the inflated uncertainties are reliable, though a few percent of sources in the tails of the distribution (mainly binaries with periods near one year) still have overestimated uncertainties.  }
    \label{fig:nss_validation}
\end{figure*}

Figure~\ref{fig:nss_validation} shows the difference between the five-parameter and 12-parameter solutions, $\Delta \varpi = \varpi_{12\, \rm par} - \varpi_{5\,\rm par}$, normalized by its total error, $\sigma_{\Delta\varpi} = \sqrt{\sigma_{\varpi_{5\,\rm par}}^2 + \sigma_{\varpi_{12\,\rm par}}^2 }$. Here $\varpi_{12\, \rm par}$ and $\sigma_{\varpi_{12\,\rm par}}$ are the parallax and its uncertainty from the \texttt{gaiadr3.nss\_two\_body\_orbit} catalog, while  $\varpi_{5\, \rm par}$ and $\sigma_{\varpi_{5\,\rm par}}$ are from the \texttt{gaiadr3.gaia\_source} catalog. Following \citet{Nagarajan2024}, we inflate $\sigma_{\varpi_{12\,\rm par}}$ by a factor of 1.4 for sources with $G>14$ and a factor of 1.7 for sources with $G\leq 14$. Our results are not sensitive to this choice, because $\sigma_{\varpi_{12\,\rm par}}\ll \sigma_{\varpi_{5\,\rm par}}$ for the vast majority of these binaries. 

In the left panel, $\sigma_{\varpi_{5\,\rm par}}$ is the reported parallax uncertainty; in the right panel, it is inflated according to Equation~\ref{eq:f}. The resulting distribution of $\Delta \varpi /\sigma_{\Delta \varpi}$ is well approximated by a Gaussian with $\sigma=1$. A few percent of binaries in the tails of the distribution with $\left|\Delta \varpi /\sigma_{\Delta \varpi} \right | > 3$ are likely to still have underestimated uncertainties; these are mainly systems with periods close to 1 year, which have more biased parallaxes than typical sources at fixed \texttt{RUWE}. Although such binaries cannot be identified based on {\it Gaia} data alone, binaries with periods between 300 and 500 d represent only about $2\%$ of the population for a \citet{Raghavan2010} period distribution and thus have only minor effects on the $\Delta \varpi /\sigma_{\Delta \varpi}$ distribution. Overall, Figure~\ref{fig:nss_validation} suggests that inflating the five-parameter parallax uncertainties according to Equation~\ref{eq:f} results in reliable uncertainties for five-parameter solutions with large \texttt{RUWE}. 

\subsection{Wide binaries in which one component has high \texttt{RUWE}}
\label{sec:wb_validation}

We next consider the parallax difference of spatially resolved wide binaries. The two components of a bound wide binary have essentially identical true parallaxes. If both components of a binary have reliable parallax uncertainties, then the distribution of uncertainty-normalized parallax errors, $\Delta \varpi / \sigma_{\Delta \varpi}$, should be a Gaussian with $\sigma\approx 1$. Here (different from Figure~\ref{fig:nss_validation}) $\Delta \varpi$ represents the difference between the five-parameter parallaxes of the two stars. If parallax errors are underestimated, the distribution will be broader. This fact was exploited by \citet{El-Badry2021} to validate the parallax uncertainties of {\it Gaia} five-parameter solutions for sources with \texttt{RUWE} $<1.4$, and by \citet{Nagarajan2024} to validate the uncertainties of 12-parameter orbital solutions. \citet{El-Badry2021} noted that their wide binary sample implied that sources with \texttt{RUWE} $>1.4$ have underestimated parallax uncertainties, but they discarded such sources rather than attempting to infer uncertainty inflation factors. 

To validate our uncertainty inflation prescription, we consider wide binaries from the \citet{El-Badry2021} catalog in which one component has \texttt{RUWE} $>1.4$. We consider pairs with separations $\theta < 4$ arcsec, for which \citet{El-Badry2021} considered any candidate companions with $\left|\Delta\varpi/\sigma_{\Delta\varpi}\right|<6$, and we only consider pairs with a $<10\%$ chance alignment probability as estimated by \citet{El-Badry2021}. 99\% of pairs in this sample have chance alignment probabilities lower than 1\%.

\begin{figure*}
    \centering
    \includegraphics[width=\textwidth]{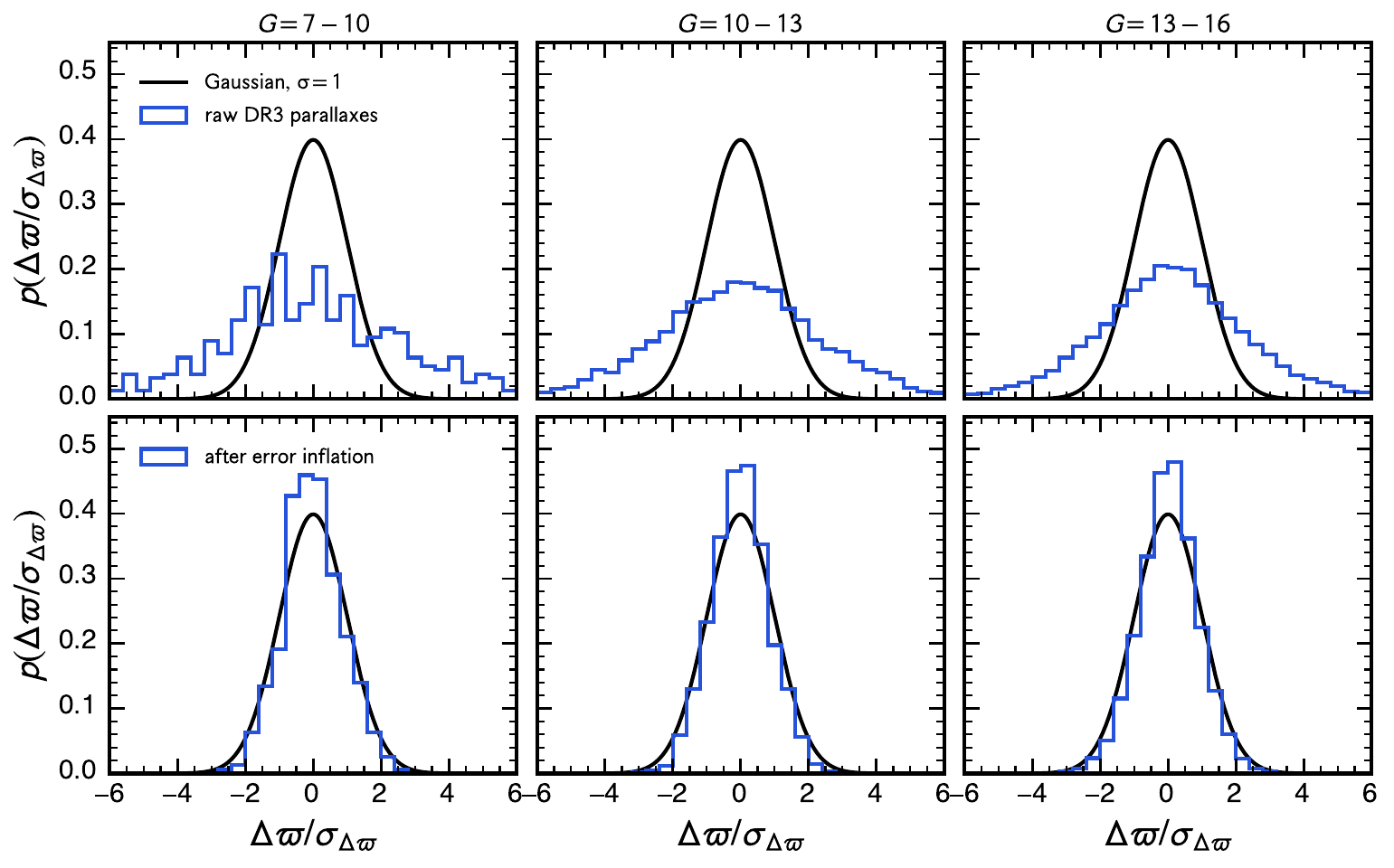}
    \caption{Uncertainty-normalized parallax difference between the two components of wide binaries in which one component has \texttt{RUWE}$>1.4$. The two components of a wide binary have essentially the same parallax, so their difference is dominated by noise, and its distribution is diagnostic of the reliability of the parallax errors. Top panels are calculated with DR3 parallax uncertainties, resulting in distributions significantly broader than the expected Gaussian with $\sigma=1$ for all magnitude ranges. Bottom panels show distributions after inflating the parallax uncertainties according to Equation~\ref{eq:f}. This results in a distribution slightly narrower than a Gaussian with $\sigma=1$, suggesting that the inflated uncertainties are realistic and that pairs in the tails of the distribution were excluded from the binary sample because their components' parallaxes were too discrepant before uncertainty inflation. }
    \label{fig:wb_validation}
\end{figure*}

Distributions of the binaries' uncertainty-normalized parallax difference before (top) and after (bottom) error inflation are shown in Figure~\ref{fig:wb_validation}. We show three bins of apparent magnitude, requiring that both components of each binary fall in the relevant bin. 

Results are similar for all three magnitude bins, and also for brighter and fainter bins not shown in the figure. Before uncertainty inflation, the distribution of $\Delta \varpi / \sigma_{\Delta \varpi}$ is significantly broader than a $\sigma=1$ Gaussian, implying uncertainties that are underestimated by a factor of $\sim2$ on average. The distributions after uncertainty inflation fully rectify this and in fact are slightly narrower than the $\sigma=1$ Gaussian. We suspect that this is a consequence of the $\left|\Delta\varpi/\sigma_{\Delta\varpi}\right|<6$ cut employed in constructing the wide binary catalog, since pairs containing components with the most strongly biased parallaxes would have been excluded from the catalog and considered unbound. We thus again conclude that Equation~\ref{eq:f} results in reliable parallax uncertainties for sources with inflated \texttt{RUWE}.

\section{Discussion and Conclusions}
\label{sec:conclusion}

Using realistic simulations of {\it Gaia} epoch astrometry, we have presented a simple empirical prescription to inflate the parallax uncertainties of {\it Gaia} sources with poor astrometric fits (Equation~\ref{eq:f} and Figure~\ref{fig:fitting_func}). This correction restores statistical consistency between expected and observed parallax offsets for binaries with known orbital solutions (Figure~\ref{fig:nss_validation}) and for components of wide binaries (Figure~\ref{fig:wb_validation}). 

In many cases, a source with high \texttt{RUWE} can still yield a useful distance constraint. For example, a solar-type star at $d=100$\,pc would have a DR3 parallax uncertainty of $\approx 0.02$ mas if it were a well-behaved single source, corresponding to a parallax SNR of $\varpi/\sigma_{\varpi}=500$. If this source had \texttt{RUWE} $= 3$, the expected formal parallax uncertainty would be $\approx 0.06$ mas, while the true uncertainty after inflation would be $\approx 0.2$ mas, still yielding a high parallax SNR of 50. On the other hand, if the same solar-type star were placed at $d=1$\,kpc, the true parallax SNR would be $\varpi/\sigma_{\varpi}\approx 33$ at \texttt{RUWE} $=1$, but only $\approx 3$ at \texttt{RUWE} $=3$.

While our fiducial simulations assume that elevated \texttt{RUWE} is always a consequence of binarity, we have verified that our inferred uncertainty inflation factors still perform well when we simply add unmodeled noise to the epoch astrometry, as might be expected if an observed source's \texttt{RUWE} is a result of photometric variability or unmodeled instrumental issues.

\subsection{Uncertainties for sources with low \texttt{RUWE}}
\citet{El-Badry2021} found that the parallax uncertainties of sources with \texttt{RUWE} $<1.4$ must be inflated by a magnitude-dependent factor to achieve statistical consistency between the parallaxes of the components of wide binaries. We find that the uncertainty inflation factors inferred by \citet{El-Badry2021} are nearly unchanged when a cut of \texttt{RUWE} $<1.1$ is used instead of \texttt{RUWE} $<1.4$ and conclude that some uncertainty inflation is still required for sources with \texttt{RUWE} $\approx 1$. That is, the inflation factors inferred by \citet{El-Badry2021} are not simply a result of sources with e.g. \texttt{RUWE} $= 1.2-1.4$. We advocate using the uncertainty inflation factors inferred here for sources with significantly inflated \texttt{RUWE}, and reverting to the prescriptions from \citet{El-Badry2021} for other sources.

A reasonable sky-averaged \texttt{RUWE} threshold for poor astrometric fits, above which uncertainties should be inflated following Equation~\ref{eq:f}, is \texttt{RUWE} $>1.25$ \citep[e.g.][]{Penoyre2022}. The distribution of \texttt{RUWE} for well-behaved sources varies somewhat across the sky, so another option is to identify sources with poor fits as those with \texttt{RUWE} significantly larger than average for their sky position and apparent magnitude. However, the origin of spatial trends in \texttt{RUWE} is not yet well understood, so caution is advisable in modeling them. We also not advocate {\it deflating} uncertainties for sources with  \texttt{RUWE} $<1$: our empirical tests with wide binaries find that such sources have similarly true uncertainties to sources with \texttt{RUWE} $\approx 1$.

\subsection{Will these prescriptions remain valid in DR4?}
\label{sec:dr4}
Our simulations assume the noise model and observing baseline of {\it Gaia} DR3, but we expect the relation between \texttt{RUWE} and parallax bias to remain quite similar in DR4 and DR5. We have verified this using \texttt{gaiamock} simulations of DR4 data, both for scenarios in which the epoch-level noise model $\sigma_{\eta}(G)$ remains the same in DR4 and scenarios in which it changes. 

As the {\it Gaia} observing baseline increases, the \texttt{RUWE} distribution for well-behaved single sources becomes more narrowly peaked at $\texttt{RUWE}=1$, while binary-driven \texttt{RUWE} enhancements become sensitive to longer periods. However, since the formal parallax uncertainties are already inflated to yield $\chi_\nu^2\approx 1$, the additional parallax inflation factor required to account for mismatch between orbital motion and a five-parameter astrometric model is primarily set by geometry and is nearly independent of the astrometric uncertainties and observing baseline at fixed \texttt{RUWE}.

\subsection{Proper motion uncertainties}
\label{sec:pms}
We have not attempted to calculate uncertainty inflation factors for proper motions. We find that the proper motion bias in DR3 is most severe for binaries with periods of $10^{3-5}$ d, somewhat longer than the observational baseline. Binaries with long periods that have negligible \texttt{RUWE} can still have highly significant proper motion bias, reflecting long-term motion of the photocenter due to the binary orbit. Proper motion bias is thus not strongly correlated with \texttt{RUWE}, though most binaries with large \texttt{RUWE} do also have underestimated proper motion uncertainties. 

The magnitude of the proper motion bias due to long-term orbital motion will not exceed the orbital velocity of the component stars, which is typically a few $\rm km\,s^{-1}$ at the periods most affected. Since this is still small compared to the typical space velocities of nearby stars, unrecognized proper motion bias is not likely to represent a major source of error for studies of Galactic structure, but it is potentially important for studies of dynamically cold stellar structures such as streams and wide binaries.

\subsection{Orbital solutions}
\label{sec:orbits}
As part of {\it Gaia} astrometric processing, a cascade of models with increasing complexity are fit to the data, including seven- and nine-parameter acceleration solutions and 12-parameter orbital solutions \citep{Halbwachs2023}. When such solutions are available, their parallax should generally be used over the parallax of the five-parameter solution \citep{Nagarajan2024}. An analog of \texttt{RUWE} for these solutions could likely also be used to inflate their uncertainties in the case of poor fits, but the uncertainty inflation factor is likely to have more complicated parameter dependencies than in the case of five-parameter solutions.

\section*{acknowledgments}
We thank the referee for helpful comments and Soumyadeep Bhattacharjee for discussions that prompted this work. This research was supported by NSF grant AST-2307232 and donations from Isaac Malsky. This work has made use of data from the European Space Agency (ESA)
mission {\it Gaia} (\url{https://www.cosmos.esa.int/gaia}), processed by
the {\it Gaia} Data Processing and Analysis Consortium (DPAC,
\url{https://www.cosmos.esa.int/web/gaia/dpac/consortium}). Funding
for the DPAC has been provided by national institutions, in particular
the institutions participating in the {\it Gaia} Multilateral Agreement.

\newpage

\newpage
%\bibliography{manuscript}

\bibliographystyle{mnras}

%\section{More stuff}
%\label{sec:appendix}

\end{document}